\documentclass[twoside,twocolumn]{article}

\usepackage{blindtext} 

\usepackage{graphicx}
\usepackage{subfigure}

\usepackage[sc]{mathpazo} 
\usepackage[T1]{fontenc} 
\usepackage{microtype} 

\usepackage[english]{babel} 

\usepackage[hmarginratio=1:1,top=32mm,columnsep=20pt]{geometry} 
\usepackage[hang, small,labelfont=bf,up,textfont=it,up]{caption} 
\usepackage{booktabs} 

\usepackage{lettrine} 

\usepackage{enumitem} 
\setlist[itemize]{noitemsep} 

\usepackage{abstract} 

\usepackage{titlesec} 
\renewcommand\thesection{\Roman{section}} 
\renewcommand\thesubsection{\roman{subsection}} 
\titleformat{\section}[block]{\large\scshape\centering}{\thesection.}{1em}{} 
\titleformat{\subsection}[block]{\large}{\thesubsection.}{1em}{} 

\usepackage{fancyhdr} 
\usepackage{titling} 

\usepackage{hyperref} 

\newcommand{\beq}{\begin{equation}}
\newcommand{\eeq}{\end{equation}}

\newcommand{\ball}{\begin{align}}
\newcommand{\eall}{\end{align}}

\newcommand{\beqar}{\begin{eqnarray}}
\newcommand{\eeqar}{\end{eqnarray}}

\newcommand{\ben}{\begin{enumerate}}
\newcommand{\een}{\end{enumerate}}

\pretitle{\begin{center}\huge\bfseries} 
\posttitle{\end{center}} 
\title{Clearing certain misconception in the common explanations of the aerodynamic
lift} 
\author{
\textsc{Navinder Singh}\thanks{Cell Phone: +919662680605; Landline: 00917926314457.},
 \textsc{K. Sasikumar Raja},  \textsc{P. Janardhan} \\[1ex] 
\normalsize Physical Research Laboratory, Ahmedabad, India. \\ 
\normalsize \href{mailto:navinder.phy@gmail.com}{navinder.phy@gmail.com} 
}
\date{\today} 
\renewcommand{\maketitlehookd}{%
\begin{abstract}
\noindent  
Air travel has become one of the most common means of
transportation. The most common question which is generally asked is: How does
an airplane gain lift? And the most common answer is via the Bernoulli
principle. It turns out that it is wrongly applied in common
explanations, and there are certain misconceptions. In
an alternative explanation the push of air from below the wing is
argued to be the lift generating force via Newton's law. There are
problems with this explanation too. In this paper we try to clear
these misconceptions, and the correct explanation, using the
Lancaster-Prandtl circulation theory, is discussed.
We argue that even the Lancaster-Prandtl theory at the zero angle of attack needs further insights. 
To this end, we put forward a theory which is applicable at zero angle of attack. 
A new length scale perpendicular to the lower surface of the wing is introduced and it turns out that the ratio of this length scale to the cord length of a wing is roughly $0.4930\pm0.09498$ for typical NACA airfoils that we analyzed. This invariance points to something fundamental. The idea of our theory is simple. The "squeezing" effect of the flow above the wing due to camber leads to an effective Venturi tube formation and leads to higher velocity over the upper surface of the wing and thereby reducing pressure according the Bernoulli  theorem and generating lift. Thus at zero angle of attack there is no need to invoke vortex and anti-vortex pair generation. In fact vortex and anti-vortex pair
generation cannot be justified. We come up with the equation for the lift coefficient at zero angle of attack:
\[C_l  = \frac{1}{c}\int_0^c dx \left(\frac{h_c^2}{(h_c - f(x))^2}-1\right).\]
Here $C_l$ is the lift coefficient, $h_c$ is our new length scale perpendicular to the
lower surface of the wing and $f(x)$ is the functional profile 
of the upper surface of the wing. 
\end{abstract}
}

\begin{document}

\maketitle

\section{Introduction}

The issue of the mechanism of the aerodynamical lift is one of the most vexed one 
\cite{Wal2007,Web1947, Wel1987, Fle1975,Tip2007, And2009, Cut1998, The1963, Pra1957, And2010}. 
The reason is the complexity of a real fluid flow over an airfoil which renders inappropriate the direct or oversimplified applicability of the standard arguments related to Bernoulli's theorem and Newton's dynamical laws.
In the next section, we consider them one by one.

\section{Wrong explanation no. 1: the standard explanation and "distance" argument}
The standard explanation of aerodynamic lift is based on an application of the Bernoulli 
theorem \cite{Cut1998}. In this explanation of the aerodynamic lift over an airfoil, it is generally argued that air flow is faster over the wing as compared to that underneath it, and this is due to the peculiar shape of an airfoil. Faster air flow leads to lower pressure over the top of a wing due to Bernoulli's theorem as compared to its lower surface, and thus aerodynamic lift is generated due to pressure difference above and below the wing. In other words it is a suction lift.  This is all fine, but the reason given for faster airflow over the top of a wing is not correct. It  is generally argued that air has to travel a longer distance on the upper surface due to more curvature of upper surface as compared to lower surface as air parcels which depart at the leading edge has to meet at the trailing edge.   

This explanation based on "distance" argument  is fundamentally flawed. Air parcels are not "living beings" that they have to meet again! In fact two air parcels which depart at the leading edge never meet again at the trailing edge, as the air parcel which flows over the top of the wing has much more speed than that which flows underneath. The upper one always reaches before the lower one. 
So, the explanation based on the above ``distance argument'' is not correct.

\section{Wrong explanation no. 2:  Newton's action-reaction}

In the literature, an alternative explanation is also found\cite{Wal2007, Tip2007}. In this explanation it is argued that it is not the Bernoulli theorem that leads to lesser pressure at the top and generate aerodynamic lift, rather it is the deflection of the air stream in the downward direction due to an angle of attack of the wing. The downward deflected air reacts back on the wing via action-reaction law of Newton and imparts upwards momentum to the wing thus lift. In other words the lift is generated due to the push of the air from below the wing. 

This explanation is also problamatic on many accounts:

\begin{enumerate}
\item If we use Newton's theory, the calculated  lift is proportional to angle of attack squared, not linearly proportional to the angle of attack as observed experimentally \cite{The1963}.

\item In reality air flow is not like bombarding bullets on the lower surface when the wing is at a finite angle of attack and deflecting bullets impart impulse in the opposite direction. Fluid flow is much more complex. The upper air stream which flows on the top of the wing is also deflected in the downward direction due the characteristic fluid motion (see Figure \ref{fig:coanda}) that is the Coanda effect \cite{Reb1966}.

\begin{figure}[!h]
\centering
\includegraphics[height=3.5cm]{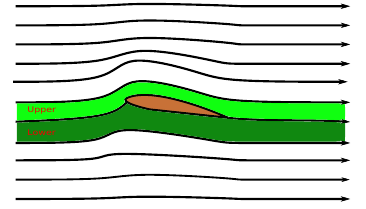}
\caption{Air is not only deflected by the lower surface 
but also by the upper surface due to Coanda effect.}
\label{fig:coanda}
\end{figure} 

\item The calculated lift slowly due to this mechanism (deflecting from the lower surface) is too less to account for the actual value of lift. In actual practice the suction lift due to Bernoulli's effect is an order of magnitude more than lift due to deflection of air from the lower surface.  

\item Also, this explanation cannot explain the finite value of lift at zero angle of attack, because then there is no deflecting surface. 

\end{enumerate}

Another  common problem is the wrong application of Newton's action-reaction. It is not the third law of Newton that is applicable here (third law is about the applied forces and reaction forces on a solid body, and their equal and opposite magnitudes). Rather it is the second law that needs to be applied consistently. The rate of change of total momentum transferred in the downward direction (of the air mass deflected by the lower surface of the wing {\em plus} the air mass deflected downwards by the upper surface due to the characteristic motion of fluid sometimes called the Coanda effect gives the upward induced lift force. And as in the case of spinning tennis ball, the calculations done by total momentum transfer method or by correct application of the Bernoulli theorem must give consistent answers. 

But the question in the Bernoulli explanation remains unanswered: why do higher speed forms on the upper surface of the wing? What is the mechanism? In the next section an answer to this question is given. 

\section{The correct explanation: the circulation theory of lift}

The correct explanation of the aerodynamic lift was given by Lancaster and Prandtl and their coworkers \cite{The1963, Pra1957}. It is based on the circulation theory of lift. In the circulation theory of lift, in addition to the laminar flow around the airfoil there is also a circulatory flow around the wing. This circulatory flow is generated due to  the formation of vortex  and anti-vortex pair. The vortex formation around the wing leads to higher air speed at the top of the wing. Then Bernoulli's theorem can be consistently applied, and mechanism of the aerodynamic lift becomes evident. The circulation theory is in good  agreement with observations \cite{The1963}. To illustrate the Lancaster-Prandtl circulation theory of lift we discuss
the following points one by one:

\begin{enumerate}

\item The mechanism of starting Vortex generation.

\item The Helmholtz theorem and Vortex - Antivortex pair generation.

\item Induced circulation around the wing profile, and the Kutta condition.

\item Induced circulation and increased speed on the top surface of the wing.

\end{enumerate}

(1) Starting Vortex generation: The most important ingredient in the circulation theory of lift is the generation of the starting vortex. In the standard literature it is only presented that with the generation of a vortex an  anti-vortex must be generated to satisfy the Helmholtz theorem \cite{The1963, Pra1957}. 
But the question remain why there should be a vortex created when a wing at a finite angle of attack is moved from rest. Here we present the actual reason behind it. 

\begin{figure}[!ht]
\begin{center}
\includegraphics[height=5.5cm]{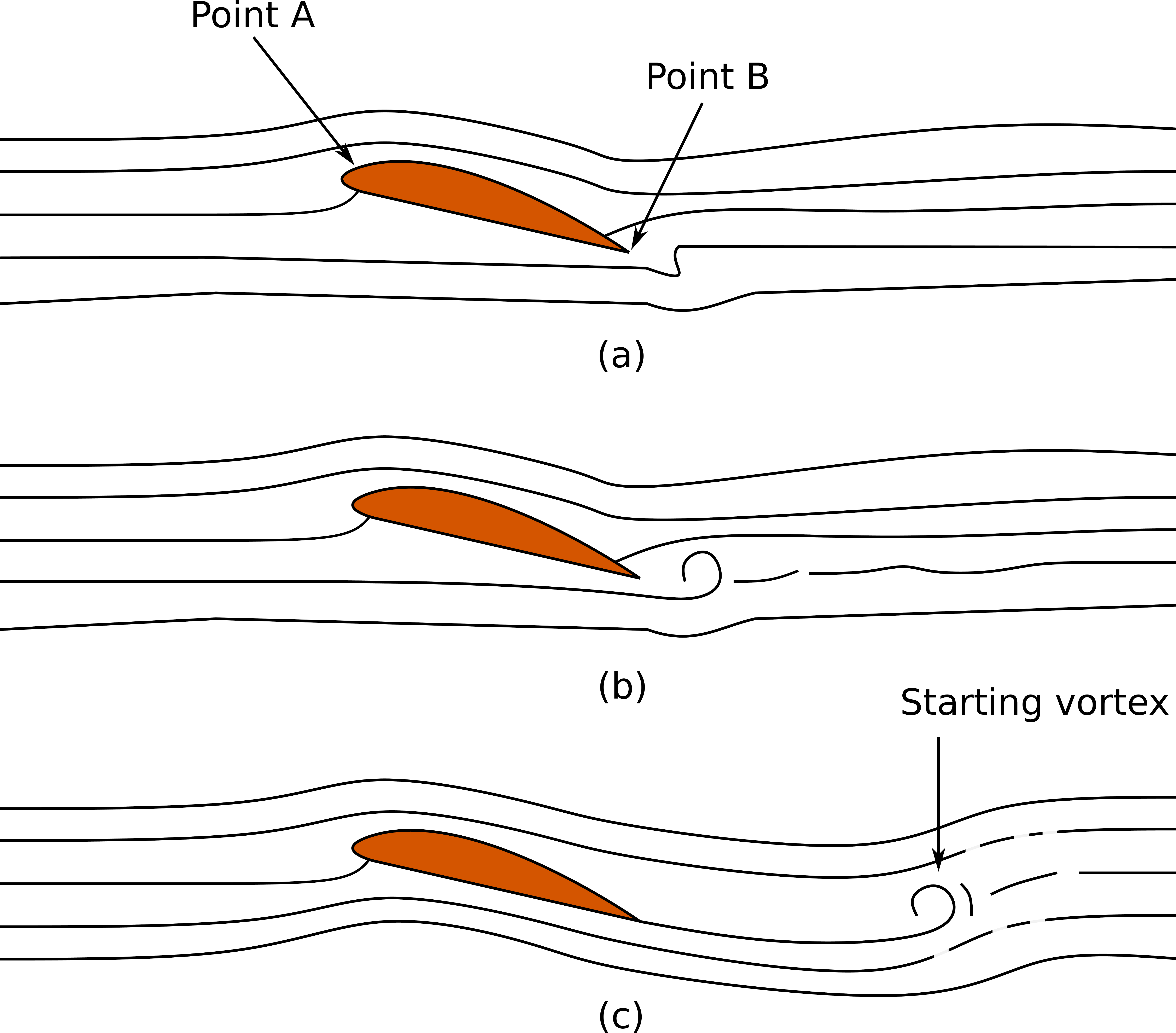}
\caption{Starting vortex generation}
\label{fig:vortex}
\end{center}
\end{figure}

When a wing is started from rest at a finite angle of attack, the air flow is little faster on the top of the leading edge due to effective Venturi formation. Similar effective venturi is formed below the trailing edge and leads to higher speed of the air stream that is leaving from below the wing (point B
in Figure \ref{fig:vortex}a). The air stream that comes from the top of the wing slows down as the effective area of cross section widens above the trailing edge (Figure \ref{fig:vortex}). 
Thus when these two air streams meet at the trailing edge, the air stream which comes from below the wing will have higher speed (this whole scenario is true only at the start of the motion of the wing from its initial rest position). Due to finite viscosity of the air, the air streams from below the wing, which are moving at higher speed tends to curl up as the stream from the upper surface has lower speed. {\em This curling up provides the seed for the formation of the starting vortex (Figure \ref{fig:vortex}) which
departs as the wing move forward}. With the generation of starting vortex we next consider the formation of vortex and anti-vortex pair. 

\begin{figure}[!ht]
\begin{center}
\includegraphics[height=2.7cm]{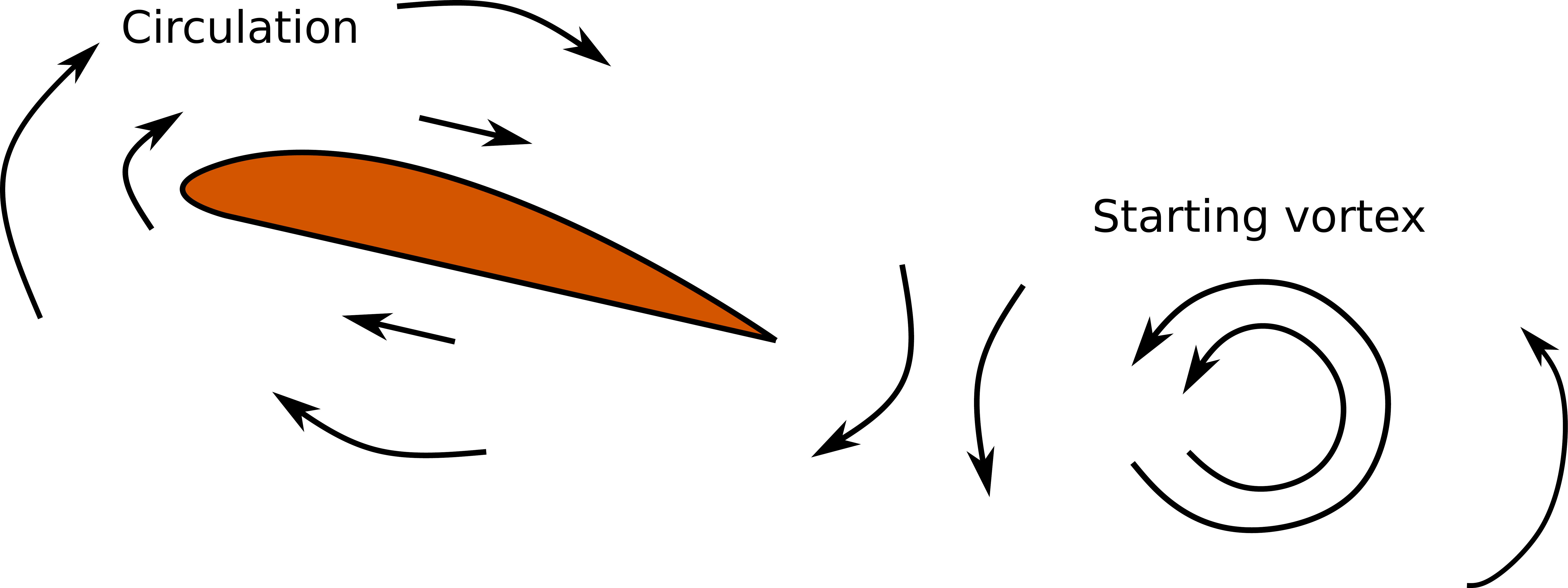}
\caption{Vortex anti-vortex pair generation due to Helmholtz theorem.}
\label{fig:pair}
\end{center}
\end{figure}

(2) The Helmholtz theorem and Vortex - Antivortex pair generation:  By Helmholtz theorem, this starting vortex cannot exist alone. The net or total circulation in  fluid must be conserved with time\cite{The1963, Pra1957}.  If there is no circulation initially in a fluid flow then a vortex generation has to be accompanied with an anti-vortex generation so that net effect is the nullification of the vortex antivortex pair. Actually, starting vortex generation due to speed difference of the two streams is the 
result of a finite viscosity\footnote{It is important to note that application of Helmholtz theorem in the present case is only an approximation, as friction is unavoidable in a real flow.}. Thus by Helmholtz theorem there is an anti-vortex generated around the wing section in response to the starting vortex (Figure \ref{fig:pair}).

(3) Induced circulation around the wing profile, and the Kutta condition: With the above picture in mind it is easy to understand the Kutta condition. In the literature Kutta condition is presented as smooth leaving of the flow from the trailing edge. But this statement hides the whole mechanism of the setting up of smooth flow at the trailing edge. And in the literature it is generally applied in the calculations of the lift coefficient without knowing why this condition has to come. At the start of a wing this condition is not justified. This condition is justified when steady state flow pattern sets up. Thus in authors' opinion it should be stated with this background made clear before hand \cite{The1963, And2010}. Initially, the flow speed of the stream coming from the lower side of the wing is higher. Then with the generation of the starting vortex, an anti-vortex is generated over the wing section. This anti vortex leads to increased speed of the stream that leaves the upper surface. It will continue to increase until the speeds of both the streams match, and flow leaves the trailing edge smoothly: the Kutta condition becomes satisfied.

\begin{figure}[!ht]
\begin{center}
\includegraphics[height=5cm]{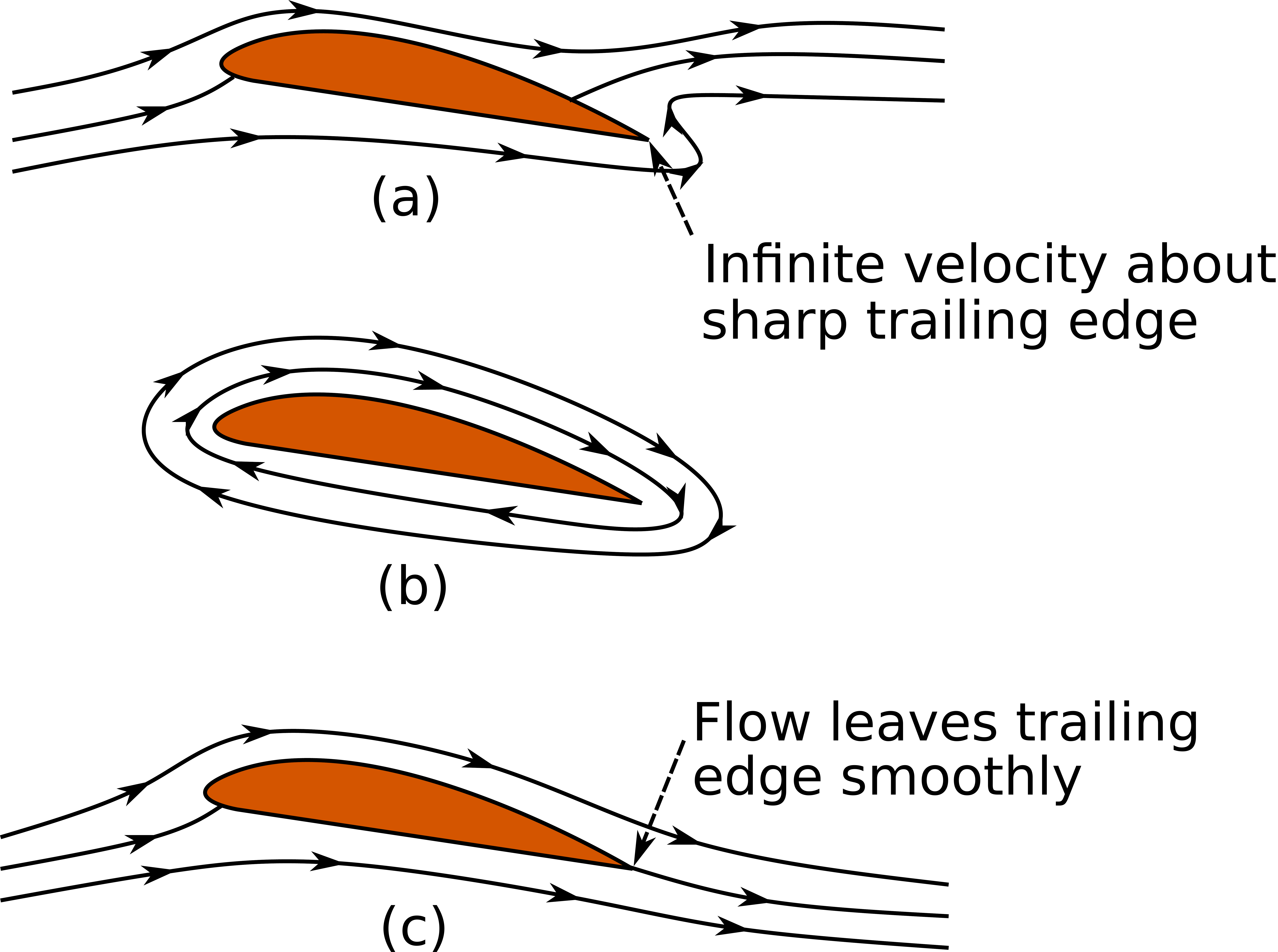}
\caption{The total flow over the top of a wing can thought of a superposition of circulatory flow and linear flow.}
\label{fig:circ}
\end{center}
\end{figure}

(4) Induced circulation and increased speed on the top surface of the wing: With this justification of the Kutta condition one obtains an increased speed of the air stream over the top of the wing span. One can resolve the total flow over the wing surface into two components: (1) stream line flow due to the motion of the wing, and (2) circulatory flow due to the generation of anti-vortex (Figure \ref{fig:circ}).

Now the stage is set for the consistent application of the Bernoulli theorem. The anti-vortex leads to higher speed over the top of the wing and lower speed comparably at the lower surface of the wing (much like in the case of spinning tennis ball). The higher speed at the top of the wing leads to lower pressure and leading to the aerodynamic lift. 

This aerodynamic lift linearly increases with the angle of attack as stronger circulation sets around the wing. The quantitative calculations agree very well with the observations \cite{The1963}.

However, at larger angle of attack, flow separation from the upper surface of the wing happens, and turbulence sets in. This leads to stall and loss of lift. The circulation theory is not competent enough to account for it. It is not theoretically possible to calculate the exact value of the angle of attack at which stall happens within the Lancaster-Prandtl theory \cite{The1963, Pra1957}.
Another problem with the circulation theory of lift is that it is not possible to explain the origin of a ``seed'' for the formation of the starting vortex at zero angle of attack. 
At zero angle of attack there is no flow squeezing below the trailing edge of the wing for a flat bottom wing, thus
the faster air stream from below the wing is not there; in other words, 
the `seed' for the starting vortex formation cannot be justified.
Due to this reason authors have developed an alternative 
explanation for the observed lift seen at zero angle of attack 
for asymmetrical airfoils. The theory is as follows.

\section{A different perspective at zero angle of attack}
In view of the above discussion, a theory of aerodynamical lift at zero angle of attack from a cambered airfoil is developed. Our theory which we call "effective Venturi tube formation" theory is different from the standard circulation theory of Lancaster and Prandtl. The idea of our theory is simple. The "squeezing" effect of the flow above the wing due to camber leads to an effective Venturi tube formation and leads to higher velocity over the upper surface of the wing and thereby reducing pressure according the Bernoulli  theorem and generating lift. 
 However, at finite angle of attack vortex generation has to be taken into account and the total lift then is given by Lancaster-Prandtl theory.
 Our theory is only valid at zero angle of attack.

\section{Formulation of the theory}
\begin{figure}[!h]
\begin{center}
\includegraphics[height=3cm]{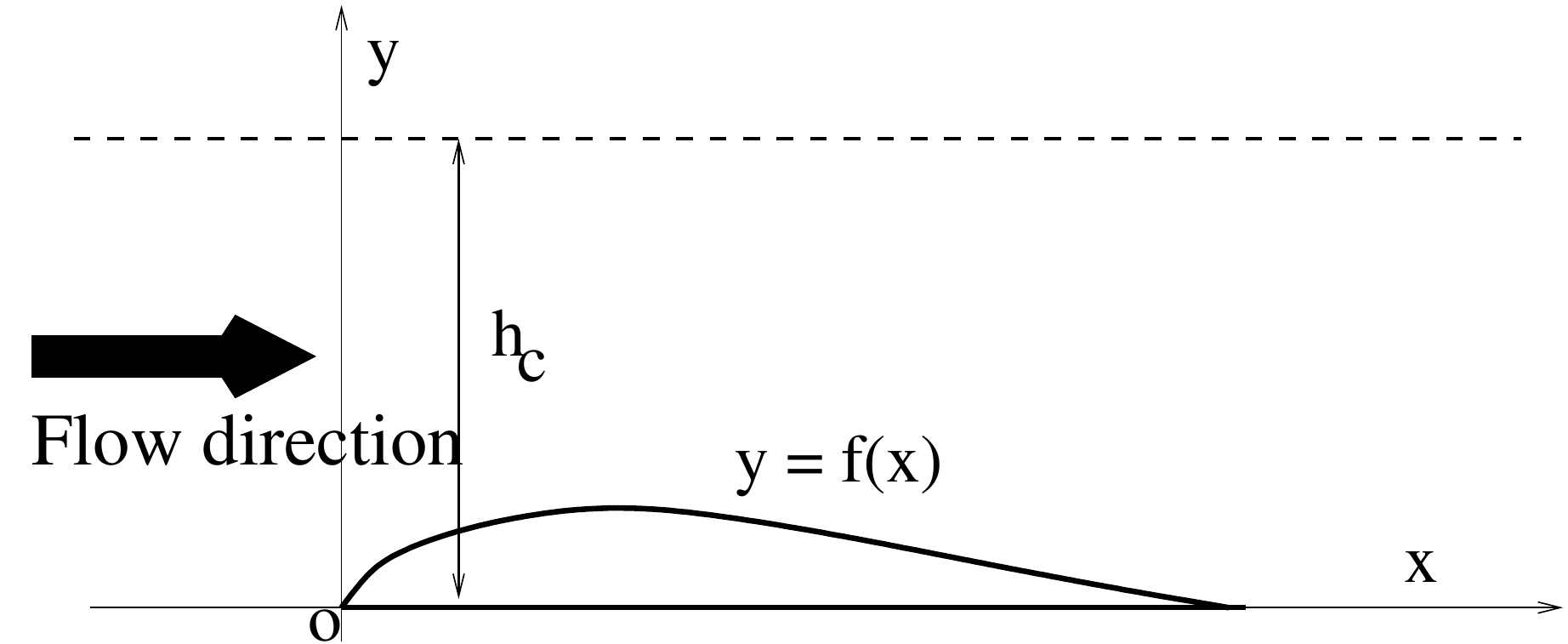}
\caption{Side view of the wing profile with various parameters defined.}
\label{fig:wing}
\end{center}
\end{figure}

Consider that we have a very long wing of cord length $c$. Let $f(x)$ be defines the curvature of the upper surface of the wing (Figure \ref{fig:wing}). Wing's leading edge is at the origin and its lower flat surface is along x-axis. Suppose that wing is moved through  air with velocity $V_\infty$ or a laminar flow is set up over a stationery wing from left with the same velocity. Imagine a height $h_c$ along the y-axis (Figure \ref{fig:wing}). 

Consider two rectangular cross-sectional areas of width $L$ . The first one (situated far in the upwind from the leading edge) has height $h_c$ (again measured from the x-axis). The other cross-sectional area is on the top of the wing with same width $L$ but with height $h_c - f(x)$ where $f(x)$ is height of the wing due to camber at position $x$. Then by continuity equation ($A_1 V_1 = A_2 V_2$) we have:
\beq
h_cLV_\infty = L (h_c-f(x))V(x)
\eeq
where $V(x)$ is the local flow velocity over the wing at position $x$. From the above equation
\beq
v(x) = \frac{h_c V_\infty}{h_c - f(x)}.
\eeq
Clearly, flow speed is increased as compared to $V_\infty$ and this leads to lower pressure according to the Bernoulli's theorem:

\beq
\frac{1}{2}\rho V_\infty^2 + P_\infty = \frac{1}{2}\rho V(x)^2 + P(x). 
\eeq
Here $\rho$ is the density of air assuming it to be incompressible and $P(x)$ is the local pressure at point $x$ ($P_\infty$ is the free stream pressure). From the above equation local pressure can be written  as

\beq
P(x) = P_\infty - \frac{1}{2}\rho (V(x)^2 - V_\infty^2).
\eeq

On substituting for $V(x)$ in the above equation and writing the reduction in pressure as $\Delta P(x) = P_\infty - P(x)$ we have

\beq
\Delta P(x) = \frac{1}{2}\rho V_\infty^2 \left(\frac{h_c^2}{(h_c - f(x))^2}-1\right).
\eeq
And the lift generated is given by $L_f = L\int_0^c dx \Delta P(x)$ with lift coefficent $C_l$ defined as

\beq
C_l = \frac{L_f}{\frac{1}{2}\rho V_\infty^2 (L c)} .
\eeq

We have 

\beq \label{eq:eq7}
C_l  = \frac{1}{c}\int_0^c dx \left(\frac{h_c^2}{(h_c - f(x))^2}-1\right).
\eeq

This is the other main result of the present contribution.

\section{Results}

\begin{table*}[!ht]
\centering
\vspace*{5px}
\begin{tabular}{|c|c|c|c|c|c|c|c|c| }
        \hline \hline

 S.&   Airfoil & Airfoil profile &$C_l$ & $h_c$\\
 No. & Model  & f(x) &  & (cm)\\
 (1) & (2) & (3) & (4) & (5) \\
    \hline
1&	2306&$-0.002920 + 0.4725x - 1.7320x^2 + 2.9018x^3 - 2.3752 x^4 + 0.7300 x^5$&0.1273&	0.50 \\
2&	2406&$0.004269 + 0.3955x - 1.1925x^2 + 1.7626x^3 - 1.3969x^4 + 0.4283x^5$	&0.1219&	0.68 \\
3&	4306&$ 0.005666 + 0.5497x - 1.8291x^2 + 3.0390x^3 - 2.6803x^4 + 0.9167x^5$	&0.2831&	0.42 \\
4&	4309&$ 0.006646+ 0.8022x - 2.8292x^2 + 4.6594x^3 - 3.8596x^4 + 1.2272x^5$	&0.2698&	0.57 \\
5&	4312&$0.0087529 + 0.9344 x - 3.5938x^2 + 6.5207 x^3 - 5.9200 x^4 + 2.0491 x^5$&0.2814&	0.59 \\
6&	4406&$ -0.001452 + 0.5682x - 1.8804 x^2 + 3.2973x^3 - 3.0847x^4 + 1.1069x^5$	&0.2960 & 0.41 \\
7&	4409&$0.004369 + 0.6216 x - 1.7009x^2 + 2.1856x^3 - 1.5131x^4 + 0.4040x^5$	&0.2682&	0.55  \\
8&	n22-il	&$0.04195 + 0.8959x - 3.5273x^2 + 6.4301x^3 - 5.8684 x^4 + 2.0331x^5$   &0.6683&	0.41 \\
9&	rhodesg32-il&$0.03741 + 0.8963x - 3.5254x^2 + 6.3863x^3 - 5.7949x^4 + 2.0015x^5$	&0.6191&	0.41  \\
10&	s7055-il&$0.006175 + 0.7466 x - 2.6465x^2 + 4.6623x^3 - 4.3012x^4 + 1.5334x^5$	&0.4095&	0.39 \\
 \hline \hline

\end{tabular}
\caption{The Table describes the various airfoil models, their profiles, experimentally measured lift coefficients and the calculated $h_c$ values using the new theory 
for zero angle of attack.}
\label{tab:one}
\end{table*}

We apply our theory to some NACA air foils with known experimental data, 
and find that the $h_c$ is more or less a
constant number. For example, with cord length $1$ cm, the functional profile of the upper surface of the airfoil model 2306 can be approximated as $f(x) = -0.002920 + 0.4725x - 1.7320x^2 + 2.9018x^3 - 2.3752 x^4 + 0.7300 x^5$. It has experimentally determined lift coefficient $C_l =  0.1273$. We numerically  solve equation (7) and determine $h_c$ for this given value of 
$C_l$ and profile $f(x)$. We find that $h_c = 0.50~\rm cm$. 
Thus there is an effective length scale at 0.50 times above the 
wing cord, above which flow can be treated undisturbed. This is reasonable intuitively. We inspected 10 airfoil models shown in column (2) of the Table \ref{tab:one}
and their corresponding upper surface profiles are listed in column (3) of the Table \ref{tab:one}. The experimentally measured lift coefficient ($C_l$) \cite{Jac1932},\footnote{\url{http://airfoiltools.com/airfoil/details?airfoil=n22-il}}, \footnote{\url{http://airfoiltools.com/airfoil/details?airfoil=rhodesg32-il}},\footnote{\url{http://airfoiltools.com/airfoil/details?airfoil=s7055-il}} and theoretically calculated $h_c$ were listed in column (4) and (5) respectively.
In our study of the 10 airfoils the mean $h_c~ =~ 0.4930$ and the standard deviation 
is $0.09498$. To test our theory we used our mean $h_c~=~0.4930$ to calculate $C_l$
for the wing section of an airfoil model `6309'. From equation \ref{eq:eq7},
we calculated $C_l$ and it is approximately 0.3891 which is close to experimentally
measured value 0.4043 for this wing section. 
 
\section{Conclusion}

In this article we have resolved various misconceptions and refuted various wrong
explanations in the issue of the mechanism of the aerodynamic lift. Actual reason for 
higher air speed on the top of the wing due to circulation mechanism is explained 
in the simple language. To that end, the physical principles of the Lancaster-Prandtl
theory are explained. We also point out that at zero angle of attack the generation of 
starting vortex and antivortex pair is questionable. The physical reason for the lack of ``seed'' formation for vortex - antivortex pair is given. To this end we have developed
a theory at zero angle of attack which seems to be in good agreement with the experimental data.

\bibliographystyle{unsrt}
\bibliography{ms}

\begin{thebibliography}{10}

\bibitem{Wal2007}
Jearl {Walker}.
\newblock {\em {The Flying Circus of Physics}}.

\bibitem{Web1947}
D.~L. {Webster}.
\newblock {What Shall We Say about Airplanes?}
\newblock {\em American Journal of Physics}, 15:228--237, May 1947.

\bibitem{Wel1987}
Klaus Weltner.
\newblock A comparison of explanations of the aerodynamic lifting force.
\newblock {\em Americal Journal of Physics}, 1987.

\bibitem{Fle1975}
N.~H. {Fletcher}.
\newblock {Mechanics of flight}.
\newblock {\em Physics Education}, 10:385--389, July 1975.

\bibitem{Tip2007}
Paul~A Tipler and Gene Mosca.
\newblock {\em Physics for scientists and engineers}.
\newblock Macmillan, 2007.

\bibitem{And2009}
{Eberhardt}~Scott {Anderson}, David~W.
\newblock {\em {Understanding flight}}.
\newblock 2009.

\bibitem{Cut1998}
JD~Cutnell and KW~Johnson.
\newblock Physics (p. 466.), 1998.

\bibitem{The1963}
Theodore Von~K{\'a}rm{\'a}n.
\newblock {\em Aerodynamics}.
\newblock Number BOOK. Cornell University press: Mc Graw-Hill company, 1963.

\bibitem{Pra1957}
Oskar Karl~Gustav Tietjens and Ludwig Prandtl.
\newblock {\em Fundamentals of hydro-and aeromechanics}, volume~1.
\newblock Courier Corporation, 1957.

\bibitem{And2010}
John~David Anderson~Jr.
\newblock {\em Fundamentals of aerodynamics}.
\newblock Tata McGraw-Hill Education, 2010.

\bibitem{Reb1966}
I.~{Reba}.
\newblock {Applications of the Coanda Effect}.
\newblock {\em Scientific American}, 214:84--92, June 1966.

\bibitem{Jac1932}
Kenneth~E. Jacobs, Eastman N.;~Ward and Robert~M. Pinkerton.
\newblock The characteristics of 78 related airfoil sections from tests in the
  variable-density wind tunnel., December 1932.

\end{thebibliography}

\end{document}